\documentclass[aps,prd,preprint,preprintnumbers,nofootinbib,superscriptaddress]{revtex4}
\usepackage{ae}
\usepackage{bm} 
\usepackage{color}
\usepackage{amsmath}
\usepackage{amssymb}
\usepackage{amsfonts}
\usepackage{graphicx}
\usepackage{slashed}
\usepackage{setspace}
\usepackage{stackrel}
\usepackage{ulem}

\newcommand{\Eqref}[1]{Eq.~\eqref{#1}}

\allowdisplaybreaks

\begin{document}

\setlength{\unitlength}{1mm}
\title{The photon polarization tensor in pulsed Hermite- and Laguerre-Gaussian beams}
\author{Felix Karbstein}\email{felix.karbstein@uni-jena.de}
\affiliation{Helmholtz-Institut Jena, Fr\"obelstieg 3, 07743 Jena, Germany}
\affiliation{Theoretisch-Physikalisches Institut, Abbe Center of Photonics, \\ Friedrich-Schiller-Universit\"at Jena, Max-Wien-Platz 1, 07743 Jena, Germany}
\author{Elena A. Mosman}\email{mosmanea@tpu.ru}
\affiliation{National Research Tomsk Polytechnic University, Lenin Ave. 30, 634050 Tomsk, Russia}
\affiliation{Physics Faculty, Tomsk State University, Lenin Ave. 36, 634050 Tomsk, Russia}

\date{\today}

\begin{abstract}
In this article, we provide analytical expressions for the photon polarization tensor in pulsed Hermite- and Laguerre-Gaussian laser beams.
Our results are based on a locally constant field approximation of the one-loop Heisenberg-Euler effective Lagrangian for quantum electrodynamics.
Hence, by construction they are limited to slowly varying electromagnetic fields, varying on spatial and temporal scales significantly larger than the Compton wavelength/time of the electron.
The latter criterion is fulfilled by all laser beams currently available in the laboratory.
Our findings will, e.g., be relevant for the study of vacuum birefringence experienced by probe photons brought into collision with a high-intensity laser pulse which can be represented as a superposition of either Hermite- or Laguerre-Gaussian modes.
\end{abstract}

\maketitle

\section{Introduction}\label{sec:Intro}

The vacuum of quantum field theory is characterized by the omnipresence of fluctuations of the theory's  particle degrees of freedom in virtual processes, describing their spontaneous creation and annihilation.
In the language of Feynman diagrams, these processes correspond to diagrams without any external lines.
As they do not couple to in- and outgoing real particles by definition, vacuum fluctuations are {\it per se} not observable.
The situation, however, changes in the presence of an external electromagnetic field.
Due to the fact that electromagnetic fields couple to charges, vacuum fluctuations of charged particles generically give rise to effective self-couplings of the prescribed electromagnetic field \cite{Euler:1935zz,Heisenberg:1935qt,Weisskopf}.

Here, we aim at studying effects of vacuum fluctuations on photon propagation at low energies in prescribed external electromagnetic fields provided by high-intensity laser pulses.
This parameter regime is governed by quantum electrodynamics (QED) subjected to an external field; cf., e.g., Ref.~\cite{Gies:2016yaa} and references therein.
Correspondingly, our quantum  particle degrees of freedom are electrons, positions and photons.
In QED, a given vacuum diagram scales as $\sim(\frac{\alpha}{\pi})^{\ell-1}$, where $\ell\in\mathbb{N}^+$ is the number of loops of the diagram,\footnote{Throughout this article, we use units where $c=\hbar=1$.} $\alpha=\frac{e^2}{4\pi}\approx\frac{1}{137}$ is the fine-structure constant, and $e$ is the electron charge.
Hence, the leading vacuum fluctuations scaling as $\sim(\frac{\alpha}{\pi})^0$ amount to single electron-positron loops, featuring $2n$, with $n\in\mathbb{N}_0$, couplings to the external electromagnetic field.
Due to Furry's theorem (charge conjugation symmetry of QED) the coupling is even in the external field.
For constant electromagnetic fields, $F^{\mu\nu}=\text{const}.$, all these diagrams can be resummed explicitly, constituting the renowned one-loop Heisenberg-Euler effective action $\Gamma_\text{HE}[{\cal F},{\cal G}^2]=V^{(4)}{\cal L}({\cal F},{\cal G}^2)$ \cite{Heisenberg:1935qt,Schwinger:1951nm}, where $V^{(4)}=\int{\rm d}^4 x$ denotes the space-time volume and ${\cal L}({\cal F},{\cal G}^2)$ is the Heisenberg-Euler effective Lagrangian. In this case, $\Gamma$ and ${\cal L}$ are trivially related, and depend on the external field only via ${\cal F}=\frac{1}{4}F_{\mu\nu}F^{\mu\nu}=\frac{1}{2}(\vec{B}^2-\vec{E}^2)$ and ${\cal G}=\frac{1}{4}F_{\mu\nu}{}^*\!F^{\mu\nu}=-\vec{E}\cdot\vec{B}$, where ${}^*\!F^{\mu\nu}=\frac{1}{2}\epsilon^{\mu\nu\alpha\beta}F_{\alpha\beta}$ is the dual field strength tensor; $\epsilon^{0123}=1$. Our metric convention is $g^{\mu\nu}=\rm{diag}(-,+,+,+)$.

As photons are massless, the only dimensionful parameter in QED is the electron/positron mass $m_e\approx511\,{\rm keV}$, setting the typical scale of the theory.
Note, that $m_e^2$ can be converted into the units of the electric and magnetic field, respectively.
Correspondingly, we have $\frac{eE}{m_e^2}\approx\frac{E[\rm{V}/\rm{m}]}{1.3\cdot10^{18}}$ and $\frac{eB}{m_e^2}\approx\frac{B[\rm{T}]}{4.4\cdot10^9}$.
These dimensionless ratios fulfill $\{\frac{eE}{m_e^2},\frac{eB}{m_e^2}\}\ll1$ for present and near-future high-intensity lasers, reaching peak field strengths up to $E\approx10^{14}$V/m and $B\approx10^6$T.
For various theoretical proposals and experimental attempts to verify effective nonlinearities of QED in external fields, we refer the reader to the pertinent reviews~\cite{Dittrich:1985yb,Fradkin:1991zq,Dittrich:2000zu,Marklund:2008gj,Dunne:2008kc,Heinzl:2008an,DiPiazza:2011tq,Dunne:2012vv,Battesti:2012hf,King:2015tba} and references therein.

A central object in the study of such effects is the photon polarization tensor, which encodes information about non-trivial modifications of the dispersion relation for probe photons propagating in the electromagnetized QED vacuum, and is the fundamental quantity in the theoretical analysis of vacuum birefringence \cite{Toll:1952,Baier}.
The one-loop photon polarization tensor is known analytically for both homogeneous electromagnetic fields \cite{BatShab,narozhnyi:1968,ritus:1972,Tsai:1974fa,Tsai:1974ap,Baier:1974hn,Urrutia:1977xb,Dittrich:2000wz,Schubert:2000yt,Dittrich:2000zu,Karbstein:2013ufa}, and generic plane wave backgrounds \cite{Baier:1975ff,Becker:1974en,Meuren:2013oya,Gies:2014jia}.
Besides, numerical results for inhomogeneous magnetic backgrounds are available from worldline Monte Carlo simulations \cite{Gies:2011he}, and analytical results for low-energy photons in slowly varying inhomogeneous electromagnetic fields were obtained in \cite{Karbstein:2015cpa}.

More specifically, the latter derivation makes use of the fact that for probe photons and background fields\footnote{If the background field is provided by a laser, as assumed here, its dominant momentum scale of variation is given by the laser frequency $\Omega$.} with frequencies and momenta delimited from above by $\upsilon\ll m_e$,
the dominant contribution to the photon polarization tensor can be inferred straightforwardly from the constant-field result in two steps \cite{BialynickaBirula:1970vy,Karbstein:2015cpa,Gies:2016yaa}:
First, the Heisenberg-Euler Lagrangian in constant fields is adopted to inhomogeneous electromagnetic fields as
\begin{equation}
 {\cal L}_\text{HE}({\cal F},{\cal G}^2)\ \xrightarrow{F^{\mu\nu}\to F^{\mu\nu}(x)}\ {\cal L}_\text{HE}\bigl({\cal F}(x),{\cal G}^2(x)\bigr)\,,
 \label{eq:LtoLx}
\end{equation}
This approximate result for ${\cal L}_\text{HE}$ differs from the -- typically unknown -- exact result in the considered inhomogeneous field by terms of ${\cal O}\bigl((\frac{\upsilon}{m_e})^2\bigr)$ \cite{Karbstein:2015cpa,Galtsov:1982}.
The effective action in inhomogeneous fields associated with the right-hand side of \Eqref{eq:LtoLx} is then defined as
\begin{equation}
 \Gamma_\text{HE}\bigl[{\cal F}(x),{\cal G}^2(x)\bigr]:=\int{\rm d}^4x\,{\cal L}_\text{HE}\bigl({\cal F}(x),{\cal G}^2(x)\bigr)\,. \label{eq:Gammax}
\end{equation}
Second, the associated photon polarization tensor in momentum space is derived from \Eqref{eq:Gammax} as usual, via \cite{Karbstein:2015cpa,Gies:2016yaa,Adorno:2017prx}\footnote{For completeness, note that this definition of the photon polarization tensor differs from that of Ref.~\cite{Karbstein:2015cpa} by an overall minus sign.}
\begin{equation}
 \Pi^{\mu\nu}(k,k')=\frac{\delta^2 \Gamma_\text{HE}\bigl[{\cal F}(x),{\cal G}^2(x)\bigr]}{\delta A_\mu(k)\,\delta A_\nu(k')}\,. \label{eq:PiDef}
\end{equation}
It is straightforward to infer that this approximate result differs from the exact expression of the photon polarization tensor in the inhomogeneous field under consideration by contributions $\sim\upsilon^2{\cal O}\bigl((\frac{\upsilon}{m_e})^2\bigr)$ \cite{Karbstein:2015cpa}.

In this article we use this approach to obtain analytical insights into the photon polarization tensor for low-energy (but not necessarily on-shell) probe photons $a^\mu(k)$, with $\{\omega,|\vec{k}|\}\ll m_e$, in linearly polarized, pulsed Hermite- and Laguerre-Gaussian laser beams of frequency $\Omega\ll m_e$ and arbitrary mode composition.
As any paraxial beam can  be expanded into either  Laguerre- or  Hermite-Gaussian
modes, our results will allow for an analytical study of signatures of vacuum nonlinearities in experimentally realistic field configurations provided by high-intensity lasers in unprecedented detail.

Our article is organized as follows: First,  in Sec.~\ref{sec:tf} we briefly recall the theoretical foundations of our approach. Second, in Sec.~\ref{sec:LG} we detail about the electromagnetic field configurations of pulsed paraxial Laguerre- and Hermite-Gaussian beams.
Section~\ref{sec:pi} is devoted to the explicit results for the corresponding photon polarization tensors.
In this context, we also discuss the limits of validity of our results, which are based upon several approximations.
Finally, we end with Conclusions and an Outlook in Sec.~\ref{sec:co}.

Besides, in App.~\ref{app:A1} we briefly detail on the overlap integrals of the field profiles of two different modes: App.~\ref{app:LG} is devoted to Laguerre- and App.~\ref{app:HG} to Hermite-Gaussian modes.
These considerations are relevant for checking the orthogonality of two given modes.
Moreover, in App.~\ref{app:A2} we use these results to provide compact formulas relating the peak field strength of a given mode to the laser pulse energy put into this mode.

\section{Theoretical foundations}\label{sec:tf}

In the paraxial approximation, the propagation direction of the laser beam is characterized by a single, globally fixed wave vector pointing along $\hat{\vec{e}}_\kappa$. 
At leading order in the diffraction angle $\theta\ll1$ (cf. Sec.~\ref{sec:LG} below) \cite{Salamin:2002dd}, the associated spatio-temporally varying electric and magnetic fields are given by $\vec{E}={\cal E}\hat{\vec{e}}_E$  and $\vec{B}={\cal E}\hat{\vec{e}}_B$, i.e., are described by the single amplitude profile $\cal E$. The unit vectors introduced here fulfill $\hat{\vec{e}}_E\cdot\hat{\vec{e}}_B=\hat{\vec{e}}_E\cdot\hat{\vec{e}}_\kappa=\hat{\vec{e}}_B\cdot\hat{\vec{e}}_\kappa=0$ as well as $\hat{\vec{e}}_E\times\hat{\vec{e}}_B=\hat{\vec{e}}_\kappa$, such that ${\cal F}={\cal G}=0$.

In the following, we will also make use of the definitions $\hat{e}^\mu_E:=(0,\hat{\vec{e}}_E)$, $\hat{e}^\mu_B:=(0,\hat{\vec{e}}_B)$ and $\hat\kappa^\mu:=(1,\hat{\vec{e}}_\kappa)$.
Note that $\hat\kappa^\mu$ provides a global reference direction with respect to which any probe photon momentum $k^\mu=(\omega,\vec{k})$ can be decomposed into parallel and perpendicular components,
\begin{equation}
 k^\mu=k_\parallel^\mu+k_\perp^\mu\,,\quad k_\parallel^\mu=(\omega,\vec{k}_\parallel)\,, \quad k_\perp^\mu=(0,\vec{k}_\perp)\,, \label{eq:k}
\end{equation}
with $\vec{k}_\parallel\equiv(\vec{k}\cdot\hat{\vec{e}}_\kappa)\hat{\vec{e}}_\kappa$ and $\vec{k}_\perp=\vec{k}-\vec{k}_\parallel$.

\begin{figure}
\center
\includegraphics[width=0.3\textwidth]{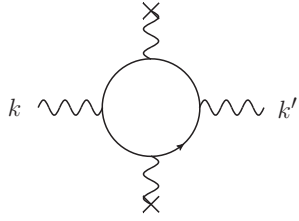}
\caption{Representative Feynman diagram for the photon polarization tensor~\eqref{eq:Picrossed}
featuring two couplings to the external field $eF^{\mu\nu}(x)$, depicted by wiggly lines ending at crosses; \Eqref{eq:Picrossed} accounts for all different possibilities to insert the external field in the fermion loop. As inhomogeneous fields $F^{\mu\nu}(x)$ can transfer energy and momentum to the fermion loop, the outgoing photon momentum $k'^\mu$ generically differs from the incident one $k^\mu$.}
\label{fig:Pi}
\end{figure}

As briefly recalled in Sec.~\ref{sec:Intro}, and detailed in Ref.~\cite{Karbstein:2015cpa}, resorting to a locally constant field approximation for the one-loop Heisenberg-Euler effective Lagrangian, the photon polarization tensor in the background field configuration introduced above is given by 
\begin{equation}
 \Pi^{\rho\sigma}(k,k') = -\frac{\alpha}{\pi}\frac{1}{45}
  \int{\rm d}^4x\,{\rm e}^{{\rm i}(k+k')x} \Bigl[ 4\,(k\hat F)^\rho  (k'\hat F)^\sigma + 7\,(k\,{}^*\!\hat F)^\rho (k'\,{}^*\!\hat F)^\sigma \Bigr]\Bigl(\frac{e{\cal E}}{m_e^2}\Bigr)^2 \,, \label{eq:Picrossed}
\end{equation}
where $\hat F^{\mu\nu} := F^{\mu\nu}/{\cal E}$ denotes the normalized field strength tensor.
Here, we employed the shorthand notations $(kF)^\mu=k_\nu F^{\nu\mu}$, $(k\,{}^*\!F)^\mu=k_\nu{}^*\!F^{\nu\mu}$, $kx=k_\mu x^\mu$, etc.
Note, that generically $\hat F^{\mu\nu}\equiv\hat F^{\mu\nu}(x)$ and ${\cal E}\equiv{\cal E}(x)$.
For a graphical representation of the photon polarization tensor~\eqref{eq:Picrossed} in terms of Feynman diagrams, cf. Fig.~\ref{fig:Pi}.

The extremely simple structure of \Eqref{eq:Picrossed} can be understood as follows \cite{Karbstein:2015cpa}:
Given that ${\cal F}={\cal G}=0$ for the field configuration considered here, non-vanishing higher powers in the field strength $F^{\mu\nu}$ necessarily involve contractions with four-momenta $k^\mu$.
The leading non-zero scalars involving the field strength are $(kF)^2$ and $(kF)(k{}^*\!F)$.
In turn, potential contributions to the photon polarization tensor proportional to ${\cal E}^{2(n+1)}$ scale as $\sim\upsilon^2(\frac{e{\cal E}}{m_e^2})^2\bigl[(\frac{e{\cal E}}{m_e^2})^2{\cal O}\bigl((\frac{\upsilon}{m_e})^2\bigr)\bigr]^n$, with $n\in\mathbb{N}_0$.
As contributions beyond ${\cal E}^2$ are $\sim\upsilon^2{\cal O}\bigl((\tfrac{\upsilon}{m_e})^2\bigr)$, they are not accounted for by the approximation adopted here.

Also note that the field configuration considered here is compatible with constant crossed \cite{narozhnyi:1968} and plane-wave fields.
In the first case we would have ${\cal E}=\text{const}.$, while in the latter case ${\cal E}\sim{\cal E}_0\cos(\hat\kappa x)$ \cite{Baier:1975ff,Becker:1974en}.
However, by means of more generic amplitude profiles ${\cal E}(x)$, this field structure can also account for features beyond plane waves, such as finite transverse beam extents and focusing effects.

For linearly polarized beams, as considered in the following, the vectors $\hat{\vec{e}}_E$ and $\hat{\vec{e}}_B$, and thus also the four-vectors in \Eqref{eq:Picrossed}, 
\begin{align}
 (k\hat F)^{\mu}&=(k\hat\kappa)\hat{e}^\mu_E-(k \hat{e}_E)\hat\kappa^\mu\,, \nonumber\\
 (k\,{}^*\!\hat F)^{\mu}&=(k\hat\kappa)\hat{e}^\mu_B-(k \hat{e}_B)\hat\kappa^\mu\,,
\end{align}
are independent of the space-time coordinate $x$, and \Eqref{eq:Picrossed} can be conveniently expressed as 
\begin{equation}
 \Pi^{\rho\sigma}(k,k') = -\frac{\alpha}{\pi}\frac{1}{45}
   \Bigl[ 4\,(k\hat F)^\rho  (k'\hat F)^\sigma + 7\,(k\,{}^*\!\hat F)^\rho (k'\,{}^*\!\hat F)^\sigma \Bigr]\Pi(k+k') \,, \label{eq:Picrossed2}
\end{equation}
with
\begin{equation}
 \Pi(k+k'):=\int{\rm d}^4x\,{\rm e}^{{\rm i}(k+k')x}\,\Bigl(\frac{e{\cal E}(x)}{m_e^2}\Bigr)^2\,. \label{eq:FT}
\end{equation}
In turn, the only nontrivial step in evaluating \Eqref{eq:Picrossed2} is to perform the four-dimensional Fourier transform~\eqref{eq:FT} of the squared amplitude profile ${\cal E}(x)$ from position to momentum space. Here, $k+k'$ corresponds to the momentum transferred from the inhomogeneous background field to the probe photon field.

So far, the approach outlined above has only been adopted for paraxial Gaussian beams prepared in the fundamental (${\rm TEM}_{00}$) mode, for which both freely propagating~\cite{Karbstein:2015cpa} and couterpropagating beams~\cite{Karbstein:2016asj} were considered.
In the present article, we provide the analogous results for the entire classes of linearly polarized paraxial Gaussian beams prepared in Hermite and Laguerre modes. Along the lines of Ref.~\cite{Karbstein:2015cpa}, these results can be straightforwardly generalized to circularly polarized beams.

To make the following considerations as transparent as possible, without loss of generality we assume the Gaussian beam to propagate in $\rm z$ direction, such that $\hat\kappa^\mu=(1,\vec{e}_{\rm z})$, and to be focused at ${\rm z}=0$.
In this case, the directions of the electric and magnetic fields can be parameterized by a single angle parameter $\phi=\text{const}.$ as $\hat{\vec{e}}_E=(\cos\phi,\sin\phi,0)$ and $\hat{\vec{e}}_B=\hat{\vec{e}}_E|_{\phi\to\phi+\frac{\pi}{2}}$.

\section{Laguerre- and Hermite-Gaussian beams}\label{sec:LG}

Coherent paraxial beams can be decomposed into an infinite sum of modes, which are typically labeled by two integer indices. Two widely used bases are Laguerre-Gaussian (LG) and Hermite-Gaussian (HG) modes \cite{Siegman,SalehTeich}. 
While the former is particularly adequate for beam profiles exhibiting a circular symmetry about the beam's propagation direction, the latter is suited for beams with Cartesian symmetry.

The amplitude profile of any given LG and HG mode can be expressed as,
\begin{equation}
 {\cal E}(x)={\mathfrak E}_0\,c({\rm x},{\rm y})\,{\rm e}^{-\frac{({\rm z}-t)^2}{(\tau/2)^2}}\,\frac{w_0}{w({\rm z})}\,{\rm e}^{-\frac{r^2}{w^2({\rm z})}}\,
 \cos\bigl(\Phi_{l,N}(x)+\varphi_0\bigr)\,, \label{eq:E}
\end{equation}
where the phase $\Phi_{l,N}(x)$ encodes the longitudinal propagation properties of the mode, and ${\mathfrak E}_0$, $c({\rm x},{\rm y})$ and $\varphi_0$ are a mode-specific peak field amplitude, transverse profile and a phase offset, respectively.
Here, $r=\sqrt{{\rm x}^2+{\rm y}^2}$ and $w({\rm z})=w_0\sqrt{1+(\frac{\rm z}{{\rm z}_R})^2}$, with Rayleigh range ${\rm z}_R$ and waist size $w_0$.
The Rayleigh range ${\rm z}_R$ is the longitudinal distance from the focus for which the beam's cross section is increased by a factor of two.
For a beam of wavelength $\lambda=\frac{2\pi}{\Omega}$, we have ${\rm z}_R=\frac{\pi w_0^2}{\lambda}$ \cite{Siegman,SalehTeich}.
The first exponential factor in \Eqref{eq:E} supplements the beam with a finite pulse duration\footnote{\label{fnt:adhoc} By this factor we augment the paraxial beam solution with a finite (Gaussian shaped) pulse duration $\tau$. This {\it ad hoc} prescription neglects contributions of ${\cal O}(\frac{1}{\tau\Omega})$. The beam solution is recovered in the limit of $\tau\to\infty$.}, and the second one describes the transverse widening of the beam along $\rm z$, when going away from its focus at ${\rm z}=0$.
Finally, the ratio $\frac{w_0}{w({\rm z})}$ describes focusing effects in the longitudinal direction, and ensures that the beam's mean energy is conserved along $\rm z$.

The fundamental Gaussian mode, minimizing the product of the focus spot size and the diffraction angle $\theta\simeq\frac{w_0}{{\rm z}_R}=\frac{2}{w_0\Omega}$, is contained in both bases. Note, that the waist size $w_0$ amounts to the focus spot radius of a beam prepared in this mode.
The paraxial approximation is valid for small values of $\theta$ and neglects terms of ${\cal O}(\theta)$.

The phase $\Phi_{l,N}(x)$ in \Eqref{eq:E} is of the following form,
\begin{equation}
    \Phi_{l,N}(x)=
    \Omega({\rm z}-t)+\frac{\Omega r^2}{2R({\rm z})}-(N+1)\arctan\Bigl(\frac{\rm z}{{\rm z}_R}\Bigr)-l\varphi\,, \label{eq:Phi}
\end{equation}
where $\varphi:=\arg({\rm x}+{\rm i}{\rm y})$ is the azimuth in cylindrical coordinates $(r,\varphi,{\rm z})$, 
$R({\rm z})={\rm z}[1+(\frac{{\rm z}_R}{{\rm z}})^2]=\frac{\Omega}{2}\frac{{\rm z}_R}{\rm z}w^2({\rm z})$ is the radius of curvature of the beam's wave fronts, and the term $\sim\arctan\bigl(\tfrac{\rm z}{{\rm z}_R}\bigr)$ accounts for the Gouy phase shift of the mode; cf., e.g., Refs.~\cite{Siegman,SalehTeich}.

More specifically, for LG modes labeled by $\{l\in\mathbb{Z},p\in\mathbb{N}_0\}$ we have
\begin{align}
 \Phi_{l,N}(x)\quad \to&\quad\ \Phi_{l,|l|+2p}(x)\,, \nonumber \\
  \varphi_0\quad \to&\quad\ \varphi_{l,p}\,, \nonumber \\
 {\mathfrak E}_0\,c({\rm x},{\rm y})\quad \to&\quad\ {\mathfrak E}_{l,p} \biggl(\frac{\sqrt{2}r}{w({\rm z})}\biggr)^{|l|} L^{|l|}_p\Bigl(\bigl(\tfrac{\sqrt{2}r}{w({\rm z})}\bigr)^2\Bigr)  \nonumber\\
 &={\mathfrak E}_{l,p} \biggl(\frac{\sqrt{2}r}{w({\rm z})}\biggr)^{|l|}\sum_{j=0}^p\frac{(-1)^j}{j!}\binom{p+|l|}{p-j}\biggl(\frac{\sqrt{2}r}{w({\rm z})}\biggr)^{2j} , \label{eq:LG}
\end{align}
where $L_p^{|l|}(\chi)$ denote generalized Laguerre polynomials and ${\mathfrak E}_{l,p}$ is the peak field amplitude of the mode.
Note, that the indices $l$ and $p$ can be identified with the mode's orbital angular momentum quantum numbers \cite{Allen:1992zz}.
To arrive at the expression in the last line of \Eqref{eq:LG}, we employed their series representation, given in formula 8.970.1 of Ref.~\cite{Gradshteyn}.
Analogously, for HG labeled by $\{m\in\mathbb{N}_0,n\in\mathbb{N}_0\}$ modes we have
\begin{align}
 \Phi_{l,N}(x)\quad \to&\quad\  \Phi_{0,m+n}(x)\,, \nonumber \\
 \varphi_0\quad \to&\quad\ \varphi_{m,n}\,, \nonumber \\
 {\mathfrak E}_0\,c({\rm x},{\rm y})\quad \to&\quad\ {\mathfrak E}_{m,n}\,H_m\bigl(\tfrac{\sqrt{2}{\rm x}}{w({\rm z})}\bigr)H_n\bigl(\tfrac{\sqrt{2}{\rm y}}{w({\rm z})}\bigr)  \nonumber\\
 &={\mathfrak E}_{m,n}\,m!n!\sum_{j=0}^{\lfloor\frac{m}{2}\rfloor}\sum_{q=0}^{\lfloor\frac{n}{2}\rfloor}\frac{2^{m+n-2(j+q)}(-1)^{j+q}}{j!q!(m-2j)!(n-2q)!}
 \biggl(\frac{\sqrt{2}{\rm x}}{w({\rm z})}\biggr)^{m-2j}\biggl(\frac{\sqrt{2}{\rm y}}{w({\rm z})}\biggr)^{n-2q} , \label{eq:HG}
\end{align}
with Hermite polynomials $H_l(\chi)$.
Here, $\lfloor n\rfloor$ is the floor function which gives as output the largest integer less than or equal to $n$.
In the last line of \Eqref{eq:HG} we made use of formula 18.5.13 of Ref.~\cite{DLMF}.
For completeness, note that the fundamental Gaussian mode amounts to the LG (HG) mode with $l=p=0$ ($m=n=0$).

\section{Photon polarization tensor in Laguerre- and Hermite-Gaussian beams}\label{sec:pi}

Before providing the explicit results for the one-loop photon polarization tensor in pulsed paraxial LG and HG beams, we briefly discuss the limits of validity of our results.
Summarizing all the points mentioned in Secs.~\ref{sec:tf} and \ref{sec:LG} above,
we find that our results neglect contributions of the following type, 
\begin{equation}
\sim\upsilon^2\Bigl[{\cal O}\bigl(\tfrac{2}{w_0\Omega}\bigr)+{\cal O}\bigl(\tfrac{1}{\tau\Omega}\bigr)+{\cal O}\bigl((\tfrac{\upsilon}{m_e})^2\bigr)+{\cal O}\bigl((\tfrac{\alpha}{\pi})^2\bigr)\Bigr]\,,
\label{eq:limitations}
\end{equation}
where $\{\omega,|\vec{k}|,\omega',|\vec{k}'|,\Omega\}\lesssim\upsilon\ll m_e$.
The overall $\upsilon^2$ dependence of \Eqref{eq:limitations} reflects the fact that the photon polarization tensor fulfills the Ward identity, $k_\rho\Pi^{\rho\sigma}(k,k')=\Pi^{\rho\sigma}(k,k')k'_\sigma=0$, implying that $\Pi^{\rho\sigma}(k,k')\sim k^\alpha k'^\beta$ \cite{Karbstein:2015cpa}.
The first term in the squared brackets arises from the restriction to the (leading order) paraxial approximation, the second one is due to our {\it ad hoc} prescription to account for a finite pulse duration, the third one refers to contributions beyond the locally constant field approximation, and the last one to contributions from higher loops.

Let us also assess the relative importance of these neglected terms.
Aiming at the experimental investigation of QED nonlinearities in external fields with lasers, high-intensity lasers and tight beam focusing are preferential, as they allow for the maximum peak field strengths in the beam focus (cf. App.~\ref{app:A2}).
Hence, we exemplarily assume the background field to be generated by a state-of-the-art high-intensity laser system, such as ELI-NP \cite{ELI}, delivering pulses of duration 
$\tau=25\,{\rm fs}\approx38.0\,{\rm eV}^{-1}$
at a wavelength of $\lambda=800\,{\rm nm}\approx4.06\,{\rm eV}^{-1}$,
corresponding to a photon energy of $\Omega=\frac{2\pi}{\lambda}\approx 1.55\,{\rm eV}$.
Moreover, we assume these pulses to be focused to $w_0=1\,\mu{\rm m}\approx5.07\,{\rm eV}^{-1}$.
For probe photons with optical to X-ray frequencies, we furthermore have $\upsilon\approx1\ldots10^4\,{\rm eV}$.
In turn, for such a scenario corrections due to the first term in \Eqref{eq:limitations} seem to be most relevant, as $\frac{2}{w_0\Omega}\approx\frac{1}{3.93}$ surpasses all the other dimensionless ratios parameterizing neglected contributions in magnitude.
\cite{Davis:1979,Barton:1989,Salamin:2002dd}.
Of course, for less tight focusing such as, e.g., $w_0=3\,\mu{\rm m}\approx15.21\,{\rm eV}^{-1}$, and thereby a substantially reduced peak intensity~\eqref{eq:Wpulse} in the beam focus, this ratio becomes smaller, and thus the paraxial approximation more justified. In the latter case we obtain $\frac{2}{w_0\Omega}\approx\frac{1}{11.79}$.

We represent our results for the one-loop polarization tensor in generic linearly polarized pulsed LG and HG beams as \Eqref{eq:Picrossed2} with
\begin{equation}
 \Pi(k+k') = \sum_{\cal N}\sum_{{\cal N}'}
 \Pi_{{\cal N};{\cal N}'}(k+k')\,, \label{eq:Pifull}
\end{equation}
where
\begin{align}
 \Pi_{{\cal N};{\cal N}'}(k+k') = \frac{e{\mathfrak E}_{\cal N}}{m_e^2}\frac{e{\mathfrak E}_{{\cal N}'}}{m_e^2}\,(2{\rm z}_R \, \pi w_0^2)\,
 \frac{\tau}{2}\sqrt{\frac{\pi}{2}}\sum_{\mathfrak{n}=-1}^{+1}{\rm e}^{-\frac{1}{8}(\frac{\tau}{2})^2(\omega+\omega'+2\mathfrak{n}\Omega)^2}\,{\cal I}_{{\cal N};{\cal N}'}^{(\mathfrak{n})}(k+k') \,, \label{eq:PiRep}
\end{align}
and the sums in \Eqref{eq:Pifull} are over all modes, i.e., ${\cal N}=\{l,p\}$ for LG and ${\cal N}=\{m,n\}$ for HG beams.
This structure is a direct consequence of the fact that \Eqref{eq:FT} is quadratic in the field strength $\cal E$ and the classical superposition principle ${\cal E}=\sum_{\cal N}{\cal E}_{\cal N}$ for linearly polarized beams.
For beams prepared in a given single mode $\cal N$, only one specific ${\cal E}_{\cal N}\neq0$ and $\Pi(k+k') =
 \Pi_{{\cal N};{\cal N}}(k+k')$.

In \Eqref{eq:PiRep} we encode the nontrivial momentum dependencies in the $\tau$ independent functions ${\cal I}_{{\cal N};{\cal N}'}^{(\mathfrak{n})}(k+k')$.
In the limit of an infinitely long pulse duration $\tau$, we have $\lim_{\tau\to\infty}\frac{\tau}{2}\sqrt{\frac{\pi}{2}}\,{\rm e}^{-\frac{1}{8}(\frac{\tau}{2})^2\phi^2} =  2\pi\,\delta(\phi)$, such that the $\mathfrak{n}=0$ term corresponds to an elastic photon scattering process with $\omega=-\omega'$,
and the $\mathfrak{n}=\pm1$ terms can be identified with inelastic processes to be associated with the absorption (emission) of two laser photons of frequency $\Omega$, i.e., $\omega=-(\omega'\pm2\Omega)$.

To keep the expressions of ${\cal I}^{(\mathfrak{n})}_{{\cal N},{\cal N}'}$ for LG and HG beams compact, we will make use of the following definition
\begin{align}
 F_\Lambda\bigl(|a|,b\bigr):&=\int_{-\infty}^\infty\frac{\rm dz}{{\rm z}_R}\,
 \Bigl(\frac{w_0}{w({\rm z})}\Bigr)^\Lambda \,{\rm e}^{-{\rm i}a\frac{\rm z}{{\rm z}_R}-b\,(\frac{w({\rm z})}{w_0})^2} \nonumber\\
 &= \delta_{0,\Lambda}\sqrt{\frac{\pi}{b}}\,{\rm e}^{-\frac{1}{b}(\frac{a}{2})^2-b} + (1-\delta_{0,\Lambda}) \frac{\sqrt{\pi}}{\Gamma(\frac{\Lambda}{2})}\int_0^\infty\frac{{\rm d}s}{s}\,\frac{s^{\frac{\Lambda}{2}}}{\sqrt{s+b}}\,{\rm e}^{-\frac{1}{s+b}(\frac{a}{2})^2-(s+b)} \,, \label{eq:F}
\end{align}
for $b\geq0$; $\delta_{0,\Lambda}$ is the Kronecker delta. As obvious from the second line of \Eqref{eq:F}, this function is well-behaved and convergent for $\Lambda\geq0$, $b\geq0$ and arbitrary values of $a$.
In the limit of $b=0$, the integration over $s$ in \Eqref{eq:F} can even be performed explicitly with the help of formula 8.432.6 of Ref.~\cite{Gradshteyn}, resulting in
\begin{equation}
 F_\Lambda\bigl(|a|,0\bigr)
 = \delta_{0,\Lambda}\, 2\pi\delta(a) + (1-\delta_{0,\Lambda})\, \frac{2\sqrt{\pi}}{\Gamma(\frac{\Lambda}{2})}\Bigl(\frac{|a|}{2}\Bigr)^{\frac{\Lambda-1}{2}}{\rm K}_{\frac{\Lambda-1}{2}}(|a|)\,,
 \label{eq:Fs}
\end{equation}
where $\Gamma(.)$ is the Gamma function, and ${\rm K}_\nu(.)$ is the modified Bessel function of the second kind.
It is possible to represent all the contributions  constituting the photon polarization tensor for LG and HG beams in terms of various parameter differentiations involving the function $F_\Lambda\bigl(|a|,b\bigr)$ introduced in \Eqref{eq:F},
implying that three out of four Fourier integrals in \Eqref{eq:Picrossed}, can be performed explicitly.

\subsection{Laguerre-Gaussian beams}

For LG beams we have ${\cal N}=\{l,p\}$, $N=|l|+2p$, and obtain
\begin{equation}
 {\cal I}_{{\cal N};{\cal N}'}^{{\rm LG}(\mathfrak{n})}(k)
 =\frac{1}{16}\sum_{j=0}^p\sum_{j'=0}^{p'}\frac{(-\sqrt{2})^{|l|+|l'|}}{ j!j'!}\binom{p+|l|}{p-j}\binom{p'+|l'|}{p'-j'} \,
 {\cal J}_{{\cal N},j;{\cal N}',j'}^{{\rm LG}(\mathfrak{n})}(k)
\end{equation}
with 
\begin{align}
  {\cal J}_{{\cal N},j;{\cal N}',j'}^{{\rm LG}(0)}(k)&=\sum_{\ell=\pm}{\rm e}^{{\rm i} \ell(\varphi_{l,p}-\varphi_{l',p'})} \bigl[1+{\rm sign}\bigl(\ell(N-N')\bigr)\partial_{h_{\rm z}}\bigr]^{|N-N'|}\, \partial_c^{j+j'} \nonumber\\ 
 &\quad\times
 \bigl({\rm i}\partial_{h_{\rm x}}+{\rm sign}(l\ell)\partial_{h_{\rm y}}\bigr)^{|l|}\bigl({\rm i}\partial_{h_{\rm x}}-{\rm sign}(l'\ell)\partial_{h_{\rm y}}\bigr)^{|l'|} \nonumber\\  
 &\quad\times \frac{1}{c}F_{|N-N'|+|l|+|l'|}\bigl(|h_{\rm z}-{\rm z}_R(k\hat\kappa)|,\tfrac{(w_0\vec{k}_\perp+\vec{h}_\perp)^2}{8c}\bigr) \Big|_{c=1,\,\vec{h}=0},
\label{eq:ILG0}
\end{align}
and
\begin{align}
 {\cal J}_{{\cal N},j;{\cal N}',j'}^{{\rm LG}(\pm 1)}(k)&={\rm e}^{\pm{\rm i}(\varphi_{l,p}+\varphi_{l',p'})}\,2^{j+j'} \bigl(1\pm\partial_{h_{\rm z}}\bigr)^{N+N'+1}\bigl(\partial_{h_{\rm x}}^2+\partial_{h_{\rm y}}^2\bigr)^{j+j'} \nonumber\\
  &\quad\times\bigl({\rm i}\partial_{h_{\rm x}}\pm{\rm sign}(l)\partial_{h_{\rm y}}\bigr)^{|l|}\bigl({\rm i}\partial_{h_{\rm x}}\pm{\rm sign}(l')\partial_{h_{\rm y}}\bigr)^{|l'|}\,{\rm e}^{-\frac{(w_0\vec{k}_\perp+\vec{h}_\perp)^2}{8}} \nonumber\\
 &\quad \times  F_{N+N'+|l|+|l'|+2(j+j'+1)}\bigl(\bigl|h_{\rm z}-{\rm z}_R(k\hat\kappa)\pm\tfrac{(w_0\vec{k}_\perp+\vec{h}_\perp)^2}{8}\bigr|,0\bigr)\Big|_{\vec{h}=0} . \label{eq:ILGpm1}
\end{align}
Here ${\rm sign}(.)$ is the sign function. With the help of \Eqref{eq:Fs}, the latter quantity can be expressed in terms of modified Bessel functions and derivatives thereof. This implies that the inelastic contributions to the photon polarization tensor in LG beams can be expressed in terms of known analytic functions for arbitrary mode numbers ${\cal N}$ and ${\cal N'}$.
Note, that for an explicit evaluation of the $\mathfrak{n}=0$ contribution with $l=l'$, the following alternative representation of \Eqref{eq:ILG0} which contains less parameter differentiations is more useful,
\begin{align}
  {\cal J}_{{\cal N},j;{\cal N}',j'}^{{\rm LG}(0)}(k)\Big|_{l=l'}&=\sum_{\ell=\pm}{\rm e}^{{\rm i} \ell(\varphi_{l,p}-\varphi_{l,p'})} \bigl[1+{\rm sign}\bigl(\ell(p-p')\bigr)\partial_{h_{\rm z}}\bigr]^{2|p-p'|}(-1)^{|l|}\, \partial_c^{|l|+j+j'} \nonumber\\ 
  &\quad\times  \frac{1}{c} F_{2|p-p'|}\bigl(|h_{\rm z}-{\rm z}_R (k\hat\kappa)|,\tfrac{(w_0\vec{k}_\perp)^2}{8c}\bigr)  \Big|_{c=1,\,h_{\rm z}=0}.
\label{eq:ILG0'}
\end{align}
Equation~\eqref{eq:F} implies that for $l=l'$ and $p=p'$, and thus ${\cal N}={\cal N}'$, \Eqref{eq:ILG0'} takes a particularly simple form: In this specific case all integrals can be performed explicitly, and we obtain
\begin{align}
  {\cal J}_{{\cal N},j;{\cal N},j'}^{{\rm LG}(0)}(k)&=(-1)^{|l|}\,\frac{4\sqrt{2\pi}}{w_0|\vec{k}_\perp|}\, \partial_c^{|l|+j+j'}\, \frac{1 }{\sqrt{c}}\,{\rm e}^{-2c\bigl(\frac{{\rm z}_R(k\hat\kappa)}{w_0|\vec{k}_\perp|}\bigr)^2-\frac{w_0^2\vec{k}_\perp^2}{8c}} \Bigg|_{c=1}.
\label{eq:ILG0"}
\end{align}

\subsection{Hermite-Gaussian beams}

Conversely, for HG beams we have ${\cal N}=\{m,n\}$, $N=m+n$, and obtain
\begin{align}
 {\cal I}_{{\cal N};{\cal N}'}^{{\rm HG}(\mathfrak{n})}(k)
 &=\frac{1}{16}\sum_{j=0}^{\lfloor\frac{m}{2}\rfloor}\sum_{q=0}^{\lfloor\frac{n}{2}\rfloor}\sum_{j'=0}^{\lfloor\frac{m'}{2}\rfloor}\sum_{q'=0}^{\lfloor\frac{n'}{2}\rfloor}\frac{m!n!m'!n'!}{j!q!j'!q'!}
 \frac{(-{\rm i})^{N+N'}\, 2^{\frac{3}{2}(N+N')-3(j+j'+q+q')}}{(m-2j)!(n-2q)!(m'-2j')!(n'-2q')!} \nonumber\\
 &\hspace{7cm}\times {\cal J}_{{\cal N},j,q;{\cal N}',j',q'}^{{\rm HG}(\mathfrak{n})}(k) ,
\end{align}
with
\begin{align}
 {\cal J}_{{\cal N},j,q;{\cal N}',j',q'}^{{\rm HG}(0)}(k)
 &= \sum_{\ell=\pm}{\rm e}^{{\rm i} \ell(\varphi_{m,n}-\varphi_{m',n'})}\bigl[1+{\rm sign}\bigl(\ell(N-N')\bigr)\partial_{h_{\rm z}}\bigr]^{|N-N'|}  \bigl(\partial_{h_{\rm x}}\bigr)^{ 2\{\frac{m+m'}{2}\} }\nonumber \\
 &\quad\times\bigl(\partial_{h_{\rm y}}\bigr)^{2\{\frac{n+n'}{2}\}}\,\Bigl(\frac{1}{2}\partial_{c_{\rm x}}\Bigr)^{\lfloor\frac{m+m'}{2}\rfloor-j-j'}\,\Bigl(\frac{1}{2}\partial_{c_{\rm y}}\Bigr)^{\lfloor\frac{n+n'}{2}\rfloor-q-q'}\frac{1}{\sqrt{c_{\rm x}c_{\rm y}}} \nonumber\\
 &\quad\times F_{|N-N'|+2\{\frac{m+m'}{2}\}+2\{\frac{n+n'}{2}\}}\bigl(|h_{\rm z}-{\rm z}_R(k\hat\kappa)|,\Sigma_{i=1}^2\tfrac{(w_0k_i+h_i)^2}{8c_i}\bigr)\Big|_{\vec{c}=1,\,\vec{h}=0} , \label{eq:IHG0}
\end{align}
where we made use of the shorthand notation $2\{\frac{n}{2}\}:=n-2\lfloor\frac{n}{2}\rfloor$, which is $0$ ($1$) for $n$ even (odd), and
\begin{align}
 \!\!{\cal J}_{{\cal N},j,q;{\cal N}',j',q'}^{{\rm HG}(\pm1)}(k)
 &={\rm e}^{\pm{\rm i}(\varphi_{m,n}+\varphi_{m',n'})}\bigl(\partial_{h_{\rm x}}\bigr)^{m+m'-2(j+j')}\,\bigl(\partial_{h_{\rm y}}\bigr)^{n+n'-2(q+q')} \bigl(1\pm\partial_{h_{\rm z}}\bigr)^{N+N'+1} \nonumber\\
 &\quad\times\!{\rm e}^{-\frac{(w_0\vec{k}_\perp+\vec{h}_\perp)^2}{8}}\! F_{2(N+N'+1-j-j'-q-q')}\!\bigl(\bigl|h_{\rm z}-{\rm z}_R(k\hat\kappa)\pm\tfrac{(w_0\vec{k}_\perp+\vec{h}_\perp)^2}{8}\bigr|,0\bigr)\Big|_{\vec{h}=0} . \label{eq:IHGpm1}
\end{align}
As for LG beams, the inelastic contributions to the polarization tensor in HG beams ${\cal J}_{{\cal N},j,q;{\cal N}',j',q'}^{{\rm HG}(\pm 1)}(k)$ can be expressed in terms of modified Bessel functions and derivatives thereof.

For the special case of a beam prepared in a distinct mode, we have $m=m'$, $n=n'$ and ${\cal N}={\cal N}'$, and obtain an expression analogous to \Eqref{eq:ILG0"} above for the elastic contribution~\eqref{eq:IHG0},
\begin{align}
 {\cal J}_{{\cal N},j,q;{\cal N},j',q'}^{{\rm HG}(0)}(k)
 &=\frac{4\sqrt{2\pi}}{w_0}\,\Bigl(\frac{1}{2}\partial_{c_{\rm x}}\Bigr)^{\lfloor\frac{m+m'}{2}\rfloor-j-j'}\,\Bigl(\frac{1}{2}\partial_{c_{\rm y}}\Bigr)^{\lfloor\frac{n+n'}{2}\rfloor-q-q'} \nonumber\\
 &\quad\times \frac{1}{\sqrt{c_{\rm y}k_{\rm x}^2+c_{\rm x}k_{\rm y}^2}}\,{\rm e}^{-2(\frac{{\rm z}_R}{w_0})^2 (k\hat\kappa)^2\bigl(\frac{k_{\rm x}^2}{8c_{\rm x}}+\frac{k_{\rm y}^2}{8c_{\rm y}}\bigr)^{-1}-w_0^2\bigl(\frac{k_{\rm x}^2}{8c_{\rm x}}+\frac{k_{\rm y}^2}{8c_{\rm y}}\bigr)}\Bigg|_{\vec{c}=1} . \label{eq:IHG0'}
\end{align}

\section{Conclusions and Outlook}\label{sec:co}

In this article we have studied the photon polarization tensor in linearly polarized, pulsed Laguerre- and Hermite-Gaussian beams propagating in the QED vacuum.
Our results are based on a locally constant field approximation of the Heisenberg-Euler effective action, and hence are manifestly limited to slowly varying electromagnetic fields which vary on scales much larger than the Compton wavelength of the electron.
This criterion is fulfilled for all current and near-future high-intensity laser fields available in the laboratory, and for probe photons with optical and X-ray frequencies $\omega$ fulfilling $\omega\ll m_e$.
The fact that we also invoke the paraxial approximation and account for a finite laser pulse duration via an {\it ad hoc} prescription, gives rise to the additional constraints of $\tau\Omega\gg1$ and $w_0\Omega\gg1$. 

As any paraxial beam can be decomposed into either Laguerre- or Hermite-Gaussian modes, our results can be considered as an important step towards the study of signatures of vacuum nonlinearities in realistically modeled high-intensity lasers fields available in experiment.
Furthermore, it is certainly interesting to study vacuum birefringence and photon diffraction effects in high-intensity laser beams prepared in a distinct higher Laguerre- or Hermite-Gaussian mode. So far, theoretical studies of these effects typically assumed the high-intensity laser beams to be prepared in the fundamental Gaussian mode \cite{Heinzl:2006xc,DiPiazza:2006pr,King:2013am,Tommasini:2010fb,King:2012aw,Dinu:2013gaa,Karbstein:2015xra,Schlenvoigt:2016,Karbstein:2016lby}; for an exception cf. Ref.~\cite{Paredes:2014oxa}.

The spatially inhomogeneous transverse intensity patterns of higher modes in the beam focus might potentially be employed to induce interference patterns in the signal photon distribution in the far field, as already theoretically analyzed and proposed as signature of vacuum nonlinearity in multi-beam configurations \cite{Hatsagortsyan:2011,King:2013am}.
The results obtained in this article will facilitate such an analysis.

\acknowledgments

The work is supported by Russian Science Foundation grant RSCF 17-72-20013.
F.K. acknowledges helpful correspondence and discussions with Holger Gies, Sebastian ``Keppi" Keppler and Matt Zepf.

\appendix

\section{Overlap integrals}\label{app:A1}

In this appendix, we briefly discuss the evaluation of the overlap integrals
of the field profiles ${\cal E}_{\cal N}(x)$ and ${\cal E}_{{\cal N}'}(x)$ of two different modes $\cal N$ and ${\cal N}'$.
More specifically, we detail on all the steps necessary for their explicit evaluation.
These integrals are, e.g., important for checking the orthogonality of two given modes, as well as for relating the peak field amplitude ${\mathfrak E}_{\cal N}$ of a given mode to the beam energy put into this mode; cf. App.~\ref{app:A2} below. 

For simplicity, we limit ourselves to the mode profiles in the beam focus at ${\rm z}=0$.
The restriction to this special case can also be justified from a physical viewpoint as the quantity of most relevance for maximizing signatures of quantum vacuum nonlinearity is the beam intensity in the focus at ${\rm z}=0$.
However, note that the respective integrations (cf. below) can also be performed analytically for ${\rm z}\neq0$ with the help of formulas 7.414.4 and 7.374.5 of \cite{Gradshteyn}. The corresponding expressions are more complicated than those for ${\rm z}=0$ and involve hypergeometric functions.

\subsection{Laguerre-Gaussian modes}\label{app:LG}

For LG beams we have ${\cal N}=\{l,p\}$, ${\cal N}'=\{l',p'\}$, and the transverse overlap integration is most conveniently performed in polar coordinates $(r,\varphi)$.
The only $\varphi$ dependence of the field profile of a LG mode is via the phase in \Eqref{eq:Phi}, which for ${\rm z}=0$ becomes
\begin{equation}
    \Phi_{l,N}(x)\big|_{{\rm z}=0}=
    -\Omega t-l\varphi\,.
\end{equation}
Hence, the integration over phase results in
\begin{align}
 &\int_0^{2\pi}{\rm d}\varphi\,\cos(-\Omega t-l\varphi+\varphi_{l,p})\,\cos(-\Omega t-l'\varphi+\varphi_{l',p'}) \nonumber\\
 &=\pi\Bigl[\delta_{l,l'}\cos(\varphi_{l,p}-\varphi_{l',p'})
 +\delta_{l,-l'}\cos(2\Omega t-\varphi_{l,p}-\varphi_{l',p'})\Bigr]\,. \label{eq:A1}
\end{align}
Equation \eqref{eq:A1} gives rise to a non-vanishing contribution only for $|l|=|l'|$.
As the field profile depends on the phase term $-l\varphi$ via the argument of a trigonometric function, which generically accounts for {\it left and right moving} components, the occurrence of contributions $\sim\delta_{l,l'}$ and $\sim\delta_{l,-l'}$ in \Eqref{eq:A1} is not surprising.
In turn, a mode labeled with $l$ is {\it a priori} not orthogonal to the one labeled with $-l$.

However, at least for an infinitely long pulse duration $\tau\to\infty$, orthogonality can be enforced by including averaging over one laser period by means of $\frac{1}{T}\int_0^{T}{\rm d}t\,(\ldots)$, with $T=\frac{2\pi}{\Omega}$, to the orthogonalization procedure.
Applying this prescription to \Eqref{eq:A1}, only the term $\sim\delta_{l,l'}$ is left, as obviously $\int_0^T{\rm d}t\,\cos(2\Omega t-\varphi_{l,p}-\varphi_{l',p'})=0$.
For finite, but slowly varying pulse durations fulfilling $\tau\Omega\gg1$ as considered here (such that the terms of ${\cal O}(\frac{1}{\tau\Omega})$ not accounted for in the adopted {\it ad hoc} prescription become negligible; cf. footnote~\ref{fnt:adhoc}), a similar prescription should still result in {\it almost orthogonality}:
In fact, if we integrate \Eqref{eq:A1} over time with the Gaussian weight accounting for the finite pulse duration in \Eqref{eq:E}, we obtain
\begin{align}
 &\int_{-\infty}^\infty{\rm d}t\,{\rm e}^{-2\frac{t^2}{(\tau/2)^2}}\int_0^{2\pi}{\rm d}\varphi\,\cos(-\Omega t-l\varphi+\varphi_{l,p})\,\cos(-\Omega t-l'\varphi+\varphi_{l',p'}) \nonumber\\
 &=\Bigl(\frac{\pi}{2}\Bigr)^{\frac{3}{2}}\tau\Bigl[\delta_{l,l'}\cos(\varphi_{l,p}-\varphi_{l',p'})
 +\delta_{l,-l'}\,{\rm e}^{-\frac{1}{8}(\tau \Omega)^2}\cos(\varphi_{l,p}+\varphi_{l',p'})\Bigr]\,, 
\intertext{and as for $\tau\Omega\gg1$ we have ${\rm e}^{-\frac{1}{8}(\tau\Omega)^2}\ll\frac{1}{\tau\Omega}$, finally}
 &=\Bigl(\frac{\pi}{2}\Bigr)^{\frac{3}{2}}\tau\Bigl[\delta_{l,l'}\cos(\varphi_{l,p}-\varphi_{l',p'})
 +{\cal O}(\tfrac{1}{\tau\Omega})\Bigr]\,. \label{eq:A2}
\end{align}

In any case, \Eqref{eq:A1} can be used to write the integral over $r$ at ${\rm z}=0$ as
\begin{align}
 &\int_0^\infty{\rm d}r\,r\, \biggl(\frac{\sqrt{2}r}{w_0}\biggr)^{2|l|}\,{\rm e}^{-2\frac{r^2}{w^2_0}}\, L^{|l|}_p\Bigl(\bigl(\tfrac{\sqrt{2}r}{w_0}\bigr)^2\Bigr)\, 
 L^{|l|}_{p'}\Bigl(\bigl(\tfrac{\sqrt{2}r}{w_0}\bigr)^2\Bigr) \nonumber\\
 &=\Bigl(\frac{w_0}{2}\Bigr)^2\int_0^\infty{\rm d}\chi\,\chi^{|l|}\,{\rm e}^{-\chi}\,L^{|l|}_p(\chi)\,L^{|l|}_{p'}(\chi)
 =\Bigl(\frac{w_0}{2}\Bigr)^2\frac{(p+|l|)!}{p!}\,\delta_{p,p'}\,,
\end{align}
where we employed table 18.3.1 of Ref.~\cite{DLMF} to perform the integral.

Putting everything together, we finally obtain
\begin{equation}
\int_{-\infty}^\infty{\rm d}t\int_0^\infty{\rm d}r\,r \int_0^{2\pi}{\rm d}\varphi\,{\cal E}_{l,p}(x){\cal E}_{l',p'}(x)\Big|_{{\rm z}=0}\approx\frac{1}{2}\Bigl(\frac{\pi}{2}\Bigr)^{\frac{3}{2}}\,\frac{(p+|l|)!}{p!}\,{\mathfrak E}_{l,p}^2\,\frac{\tau}{2}\,w_0^2\,\delta_{l,l'}\delta_{p,p'}\,,
\label{eq:A3}
\end{equation}
where we explicitly neglected terms of ${\cal O}(\frac{1}{\tau\Omega})$, which are of the same order as those neglected by the adopted {\it ad hoc} prescription to account for a finite pulse duration (cf. footnote~\ref{fnt:adhoc}).

\subsection{Hermite-Gaussian modes}\label{app:HG}

In the case of HG beams we have ${\cal N}=\{m,n\}$ and ${\cal N}'=\{m',n'\}$, and 
the transverse overlap integral is most conveniently performed in Cartesian coordinates $({\rm x},{\rm y})$.
Due to the symmetry of $c({\rm x},{\rm y})$ in \Eqref{eq:HG} under the simultaneous exchange of ${\rm x}\leftrightarrow{\rm y}$ and $m\leftrightarrow n$, it is sufficient to explicitly evaluate the integral over ${\rm x}$.
With the help of table 18.3.1 of Ref.~\cite{DLMF}, we obtain
\begin{align}
 &\int_{-\infty}^\infty{\rm dx}\,{\rm e}^{-2\frac{{\rm x}^2}{w^2_0}}\,H_m\bigl(\tfrac{\sqrt{2}{\rm x}}{w_0}\bigr)\,H_{m'}\bigl(\tfrac{\sqrt{2}{\rm x}}{w_0}\bigr) \nonumber\\
 &=\frac{w_0}{\sqrt{2}}\int_{-\infty}^\infty{\rm d}\chi\,{\rm e}^{-\chi^2}\,H_m(\chi)\,H_{m'}(\chi)
 =\sqrt{\frac{\pi}{2}}\,w_0\,2^m\,m!\,\delta_{m,m'}\,.
\end{align}
Hence, two distinct HG modes are generically orthogonal to each other under integration over the transverse coordinates.

Aiming at an integration over time, we can use the same steps as invoked in App.~\ref{app:LG} above to obtain
\begin{equation}
\int_{-\infty}^\infty{\rm d}t\,{\rm e}^{-2\frac{t^2}{(\tau/2)^2}}\,\cos^2(\Omega t)
=\sqrt{\frac{\pi}{2}}\,\frac{\tau}{4}\,\Bigl[1+{\cal O}(\tfrac{1}{\tau\Omega})\Bigr]\,,
\end{equation}
such that
\begin{equation}
\int_{-\infty}^\infty{\rm d}t \int_{-\infty}^\infty{\rm dx} \int_{-\infty}^\infty{\rm dy}\,{\cal E}_{m,n}(x){\cal E}_{m',n'}(x)\Big|_{{\rm z}=0}\approx\frac{1}{2}\Bigl(\frac{\pi}{2}\Bigr)^{\frac{3}{2}}\,2^{m+n}\,m!\,n!\,{\mathfrak E}_{m,n}^2\,\frac{\tau}{2}\,w_0^2\,\delta_{m,m'}\delta_{n,n'}\,.
\label{eq:A4}
\end{equation}

\section{Relation of peak field strength and laser pulse energy}\label{app:A2}

As $I_{\cal N}={\cal E}_{\cal N}^2$ is the intensity associated with the electromagnetic field in mode ${\cal N}$, for ${\cal N}={\cal N}'$ the expressions in Eqs.~\eqref{eq:A3} and \eqref{eq:A4} amount to the energy $W_{\cal N}$ put in mode $\cal N$, i.e.,
\begin{equation}
W_{\cal N}\approx\frac{1}{2}\Bigl(\frac{\pi}{2}\Bigr)^{\frac{3}{2}}\,{\mathfrak c}_{\cal N}\,{\mathfrak E}_{\cal N}^2\,\frac{\tau}{2}\,w_0^2
\quad\leftrightarrow\quad
{\mathfrak E}_{\cal N}^2\approx8\sqrt{\frac{2}{\pi}}\frac{1}{{\mathfrak c}_{\cal N}}\frac{W_{\cal N}}{\pi w_0^2\tau}\,,
\label{eq:Wpulse}
\end{equation}
with mode specific coefficients ${\mathfrak c}_{\cal N}$ given by
\begin{align}
 &\text{for\ LG\ modes}\ {\cal N}=\{p,l\}:\quad\ \ {\mathfrak c}_{p,l}=\frac{(p+|l|)!}{p!}\quad\quad\text{and}
 \nonumber\\
 &\text{for\ HG\ modes}\ {\cal N}=\{m,n\}:\quad{\mathfrak c}_{m,n}=2^{m+n}\,m!\,n!\,.
 \label{eq:cs}
\end{align}

Using the orthogonality of the field profiles of two distinct modes, the total laser pulse energy $W$ to be partitioned into the different modes $\cal N$ is given by 
\begin{equation}
 W=\sum_{\cal N} W_{\cal N}\,.
\end{equation}
Equations~\eqref{eq:Wpulse} and \eqref{eq:cs} imply that -- as to be expected -- when putting the total energy into a given single mode $\cal N$, i.e., $W=W_{\cal N}$, the maximum peak field strength ${\mathfrak E}_{\cal N}$ and intensity is reached in the fundamental Gaussian mode ${\cal N}=\{0,0\}$. The coefficients ${\mathfrak c}_{\cal N}$ grow with increasing $p$ and $|l|$, and $m$ and $n$, respectively.


\begin{thebibliography}{10}\setlength{\itemsep}{-0.5mm}

\bibitem{Euler:1935zz} 
  H.~Euler and B.~Kockel,
  Naturwiss.\  {\bf 23}, 246 (1935).

\bibitem{Heisenberg:1935qt} 
  W.~Heisenberg and H.~Euler,
  Z.\ Phys.\  {\bf 98}, 714 (1936),
  an English translation is available at [physics/0605038].

\bibitem{Weisskopf}
V.~Weisskopf, 
Kong.\ Dans.\ Vid.\ Selsk., Mat.-fys.\ Medd.\ {\bf XIV}, 6 (1936).

\bibitem{Gies:2016yaa} 
  H.~Gies and F.~Karbstein,
  JHEP {\bf 1703}, 108 (2017)
  [arXiv:1612.07251 [hep-th]].

\bibitem{Schwinger:1951nm} 
  J.~S.~Schwinger,
  Phys.\ Rev.\ {\bf 82}, 664 (1951).

\bibitem{Dittrich:1985yb} 
  W.~Dittrich and M.~Reuter,
  Lect.\ Notes Phys.\  {\bf 220}, 1 (1985).

\bibitem{Fradkin:1991zq} 
  E.~S.~Fradkin, D.~M.~Gitman and S.~M.~Shvartsman,
  Springer Series in Nuclear and Particle Physics (1991).

\bibitem{Dittrich:2000zu} 
  W.~Dittrich and H.~Gies,
  Springer Tracts Mod.\ Phys.\  {\bf 166}, 1 (2000).
  
\bibitem{Marklund:2008gj} 
  M.~Marklund and J.~Lundin,
  Eur.\ Phys.\ J.\ D {\bf 55}, 319 (2009)
  [arXiv:0812.3087 [hep-th]].

\bibitem{Dunne:2008kc} 
  G.~V.~Dunne,
  Eur.\ Phys.\ J.\ D {\bf 55}, 327 (2009)
  [arXiv:0812.3163 [hep-th]].

\bibitem{Heinzl:2008an} 
  T.~Heinzl and A.~Ilderton,
  Eur.\ Phys.\ J.\ D {\bf 55}, 359 (2009)
  [arXiv:0811.1960 [hep-ph]].
  
\bibitem{DiPiazza:2011tq} 
  A.~Di Piazza, C.~Muller, K.~Z.~Hatsagortsyan and C.~H.~Keitel,
  Rev.\ Mod.\ Phys.\  {\bf 84}, 1177 (2012)
  [arXiv:1111.3886 [hep-ph]].

\bibitem{Dunne:2012vv} 
  G.~V.~Dunne,
  Int.\ J.\ Mod.\ Phys.\ A {\bf 27}, 1260004 (2012) [Int.\ J.\ Mod.\ Phys.\ Conf.\ Ser.\  {\bf 14}, 42 (2012)]
  [arXiv:1202.1557 [hep-th]].

\bibitem{Battesti:2012hf} 
  R.~Battesti and C.~Rizzo,
  Rept.\ Prog.\ Phys.\  {\bf 76}, 016401 (2013)
  [arXiv:1211.1933 [physics.optics]].

\bibitem{King:2015tba} 
  B.~King and T.~Heinzl,
  High Power Laser Science and Engineering, 4, e5 (2016)
  [arXiv:1510.08456 [hep-ph]].

\bibitem{Toll:1952}
J.~S.~Toll,
Ph.D. thesis, Princeton Univ., 1952 (unpublished).

\bibitem{Baier}
R.~Baier and P.~Breitenlohner, 
{Act.~Phys.~Austriaca} {\bf 25}, 212 (1967); 
{Nuov.~Cim.~B}\ {\bf 47} 117 (1967).
  
\bibitem{BatShab}
  I.~A.~Batalin and A.~E.~Shabad,
  Zh.\ Eksp.\ Teor.\ Fiz.\  {\bf 60}, 894 (1971)
  [Sov.\ Phys.\ JETP\ {\bf 33}, 483 (1971)].

\bibitem{narozhnyi:1968}
  N.~B.~Narozhnyi,
 Zh.\ Eksp.\ Teor.\ Fiz.\ {\bf 55}, 714 (1968)
 [Sov.\ Phys.\ JETP \textbf{28}, 371 (1969)].
  
\bibitem{ritus:1972}
 V.~I.~Ritus,
 Ann.\ Phys.\ {\bf 69}, 555 (1972).

\bibitem{Tsai:1974fa}
  W.~y.~Tsai and T.~Erber,
  {\it Phys.\ Rev.\  D}\ {\bf 10}, 492 (1974).
 
\bibitem{Tsai:1974ap}
  W.~y.~Tsai,
  Phys.\ Rev.\  D {\bf 10}, 2699 (1974).
 
\bibitem{Baier:1974hn} 
  V.~N.~Baier, V.~M.~Katkov and V.~M.~Strakhovenko,
  Zh.\ Eksp.\ Teor.\ Fiz.\  {\bf 68}, 405 (1975)
  [Sov.\ Phys.\ JETP\ {\bf 41}, 198 (1975)].
 
\bibitem{Urrutia:1977xb}
  L.~F.~Urrutia,
  Phys.\ Rev.\  D\ {\bf 17}, 1977 (1978).

\bibitem{Dittrich:2000wz} 
  W.~Dittrich and R.~Shaisultanov,
  Phys.\ Rev.\ D {\bf 62}, 045024 (2000)
  [hep-th/0001171].

\bibitem{Schubert:2000yt} 
  C.~Schubert,
  Nucl.\ Phys.\ B {\bf 585}, 407 (2000)
  [hep-ph/0001288].

\bibitem{Karbstein:2013ufa} 
  F.~Karbstein,
  Phys.\ Rev.\ D {\bf 88}, 085033 (2013)
  [arXiv:1308.6184 [hep-th]].

\bibitem{Baier:1975ff} 
  V.~N.~Baier, A.~I.~Milshtein and V.~M.~Strakhovenko,
  Zh.\ Eksp.\ Teor.\ Fiz.\  {\bf 69}, 1893 (1975)
  [Sov.\ Phys.\ JETP\ {\bf 42}, 961 (1976)].

\bibitem{Becker:1974en} 
  W.~Becker and H.~Mitter,
   J.\ Phys.\ A:\ Math.\ Gen.\ {\bf 8} 1638 (1975).

\bibitem{Meuren:2013oya} 
  S.~Meuren, C.~H.~Keitel and A.~Di Piazza,
  Phys.\ Rev.\ D {\bf 88}, 013007 (2013)
  [arXiv:1304.7672 [hep-ph]].
  
\bibitem{Gies:2014jia} 
  H.~Gies, F.~Karbstein and R.~Shaisultanov,
  Phys.\ Rev.\ D {\bf 90}, 033007 (2014)
  [arXiv:1406.2972 [hep-ph]].

\bibitem{Gies:2011he} 
  H.~Gies and L.~Roessler,
  Phys.\ Rev.\ D {\bf 84}, 065035 (2011)
  [arXiv:1107.0286 [hep-ph]].

\bibitem{Karbstein:2015cpa} 
  F.~Karbstein and R.~Shaisultanov,
  Phys.\ Rev.\ D {\bf 91}, 085027 (2015)
  [arXiv:1503.00532 [hep-ph]].

\bibitem{BialynickaBirula:1970vy}
  Z.~Bialynicka-Birula and I.~Bialynicki-Birula,
  Phys.\ Rev.\ D {\bf 2}, 2341 (1970).

\bibitem{Galtsov:1982} 
  G.~V.~Galtsov and N.~S.~Nikitina,
  Zh.\ Eksp.\ Teor.\ Fiz.\  {\bf 84}, 1217 (1983) [Sov.\ Phys.\ JETP\ {\bf 57}, 705 (1983)].

\bibitem{Adorno:2017prx} 
  T.~C.~Adorno, D.~M.~Gitman and A.~E.~Shabad,
  arXiv:1710.00138 [hep-th].
  
\bibitem{Salamin:2002dd} 
  Y.~I.~Salamin, G.~R.~Mocken and C.~H.~Keitel,
  Phys.\ Rev.\ ST Accel.\ Beams {\bf 5}, 101301 (2002);
  Y.~I.~Salamin, S.~X.~Hu, K.~Z.~Hatsagortsyan and C.~H.~Keitel,
  Phys.\ Rept.\  {\bf 427}, 41 (2006).

\bibitem{Karbstein:2016asj} 
  F.~Karbstein,
  Russ.\ Phys.\ J.\  {\bf 59}, 1761 (2017)
  [arXiv:1607.01546 [hep-ph]].

\bibitem{Siegman}
A.~E.~Siegman, \textit{Lasers}, First Edition, University Science Books, USA (1986).

\bibitem{SalehTeich}
B.~E.~A.~Saleh and M.~C.~Teich, \textit{Fundamentals of Photonics}, First Edition, John Wiley \& Sons, USA (1991).

\bibitem{Allen:1992zz} 
  L.~Allen, M.~W.~Beijersbergen, R.~J.~C.~Spreeuw and J.~P.~Woerdman,
  Phys.\ Rev.\ A {\bf 45}, 8185 (1992).

\bibitem{Gradshteyn}
I.~S.~Gradshteyn and I.~M.~Ryzhik, \textit{Table of Integrals, Series, and Products}, Fifth Edition, Academic Press, UK (1994).

\bibitem{DLMF}
F.~W.~J. Olver, A.~B. {Olde Daalhuis}, D.~W. Lozier, B.~I. Schneider, R.~F. Boisvert, C.~W. Clark, B.~R. Miller and B.~V. Saunders, eds.,
{\it NIST Digital Library of Mathematical Functions}, http://dlmf.nist.gov/, Release 1.0.14 of 2016-12-21.

\bibitem{ELI}
See the ELI website: 
https://eli-laser.eu/ (2017).

\bibitem{Davis:1979} 
  L.~W.~Davis,
  Phys.\ Rev.\ A {\bf 19}, 3 (1979).
  
\bibitem{Barton:1989}
 J.~P.~Barton and D.~R.~Alexander,
 J. Appl. Phy. {\bf 66}, 2800 (1989).

\bibitem{Heinzl:2006xc} 
  T.~Heinzl, B.~Liesfeld, K.~U.~Amthor, H.~Schwoerer, R.~Sauerbrey and A.~Wipf,
  Opt.\ Commun.\  {\bf 267}, 318 (2006).

\bibitem{DiPiazza:2006pr} 
  A.~Di Piazza, K.~Z.~Hatsagortsyan and C.~H.~Keitel,
  Phys.\ Rev.\ Lett.\  {\bf 97}, 083603 (2006)
  [hep-ph/0602039].

\bibitem{King:2013am} 
  B.~King, A.~Di Piazza and C.~H.~Keitel,
 Nature Photon.\ {\bf 4}, 92 (2010)
 [arXiv:1301.7038 [physics.optics]]; 
 Phys.\ Rev.\ A {\bf 82}, 032114 (2010)
 [arXiv:1301.7008 [physics.optics]].

\bibitem{Tommasini:2010fb} 
  D.~Tommasini and H.~Michinel,
  Phys.\ Rev.\ A {\bf 82}, 011803 (2010)
  [arXiv:1003.5932 [hep-ph]].

\bibitem{King:2012aw} 
  B.~King and C.~H.~Keitel,
  New J.\ Phys.\  {\bf 14}, 103002 (2012)
  [arXiv:1202.3339 [hep-ph]].

\bibitem{Dinu:2013gaa} 
  V.~Dinu, T.~Heinzl, A.~Ilderton, M.~Marklund and G.~Torgrimsson,
  Phys.\ Rev.\ D {\bf 89}, 125003 (2014)
  [arXiv:1312.6419 [hep-ph]]; 
  Phys.\ Rev.\ D {\bf 90}, 045025 (2014)
  [arXiv:1405.7291 [hep-ph]].
 
\bibitem{Karbstein:2015xra} 
  F.~Karbstein, H.~Gies, M.~Reuter and M.~Zepf,
  Phys.\ Rev.\ D {\bf 92}, 071301 (2015)
  [arXiv:1507.01084 [hep-ph]].
  
\bibitem{Schlenvoigt:2016} 
 H.~-P.~Schlenvoigt, T.~Heinzl, U.~Schramm, T.~Cowan and R.~Sauerbrey,
 Physica\ Scripta {\bf 91}, 023010 (2016).
  
\bibitem{Karbstein:2016lby} 
  F.~Karbstein and C.~Sundqvist,
  Phys.\ Rev.\ D {\bf 94}, 013004 (2016)
  [arXiv:1605.09294 [hep-ph]].

\bibitem{Paredes:2014oxa} 
  A.~Paredes, D.~Novoa and D.~Tommasini,
  Phys.\ Rev.\ A {\bf 90}, 063803 (2014)
  [arXiv:1412.3390 [physics.optics]].

\bibitem{Hatsagortsyan:2011}
K.~Z.~Hatsagortsyan and G.~Y.~Kryuchkyan,
 Phys.\ Rev.\ Lett. {\bf 107}, 053604 (2011).
 
\end{thebibliography}
\end{document}